# Tunable Mid-Infrared Chiral Selective Absorber Based on Asymmetric V-Shaped Metasurfaces Driven by Chiral quasi-Bound State in the Continum

Yuqing Liu, Yi Wang, Ruihan Ma, Nianzhao Wang, Mengtao Sun[1] and Yuqing Cheng[2]

School of Mathematics and Physics, University of Science and Technology Beijing, Beijing 100083, People's Republic of China

**Abstract:** Efficient discrimination of circularly polarized light (CPL) is of great significance in photonics. In this study, we propose a nanophotonic device based on asymmetric V-shaped metasurfaces that exhibits strong discrimination between left- and right-circularly polarized (LCP and RCP) light at the resonant wavelengths of the absorption spectra. The chiral-selective response originates from a quasi-bound state in the continuum (qBIC) mode enabled by controlled symmetry breaking in the V-shaped structure, which exhibits high absorption under LCP illumination while the resonance is strongly suppressed under RCP incidence, leading to a large absorption contrast between the two circular polarizations. This significant absorption difference enables highly efficient chiral discrimination. Furthermore, the resonant wavelength can be broadly tuned from 5200 to 6200 nm by scaling the structural dimensions without compromising the high absorption contrast between LCP and RCP. This work not only enables high-performance chiral detection and separation, but also offers valuable guidance for reconfigurable chiral nanodevices, with promising applications in areas such as enantiomer sensing, chiral imaging, and CPL spectroscopy.

## 1. Introduction

Precise discrimination of circularly polarized light (CPL) is crucial in modern photonics, with broad applications in chiral molecule sensing, optical communications, and quantum information processing[1-5]. In sensing, amplifying the weak absorption difference of chiral molecules to CPL via nanostructures is the core of ultra-sensitive chiral detection[6-9]. In communications, separating orthogonal spin states is the foundation for on-chip routing and spin multiplexing[10-13]. Therefore, developing devices for efficient CPL discrimination is of great significance for advancing sensing and integrated photonics[14-16].

[1] Email: mengtaosun@ustb.edu.cn
[2] Email: yuqingcheng@ustb.edu.cn

Recently, quasi-bound states in the continuum (qBICs) have emerged as promising resonant modes for chiral photonic devices. Originally, bound states in the continuum (BICs) are non-radiative states embedded within the radiation continuum that cannot be excited by free-space light due to symmetry-protected radiation cancellation. Introducing structural symmetry breaking disrupts this cancellation and converts BICs into qBICs-leaky resonances with finite yet controllable radiation linewidths and strong near-field enhancement. Notably, when the symmetry breaking itself carries chiral character, the resulting chiral qBICs exhibit intrinsic photonic spin selectivity, enabling selective excitation by specific circularly polarized light and spin-dependent absorption. The theoretical foundation for chiral qBICs in planar metasurfaces was first established by Overvig et al.[17] who demonstrated that symmetry-broken structures can support qBIC resonances with tailored elliptical polarization states, thereby extending the qBIC concept from linear to arbitrary elliptical polarization and enabling geometric phase engineering for spin-selective light–matter interactions. Subsequent experimental studies have realized chiral qBIC–driven perfect absorption and high-contrast CPL discrimination in practical metasurface systems[18].

In recent years, researchers have developed a variety of distinctive methods for CPL detection, which can be roughly divided into two classes: devices relying on geometrically symmetric structures that exploit extrinsic physical mechanisms and <devices based on geometrically asymmetric or intrinsically chiral materials[10, 14, 19]. For instance, Zhang et al.[20] utilized symmetric V-shaped groove arrays in achiral dielectric nanostructures to directly discriminate between left- and right-circularly polarized (LCP and RCP) light via the photo-thermoelectric effect, achieving an unprecedented circular polarization discrimination ratio of up to 107 in the visible regime. Distinct from this symmetric-geometry approach, asymmetric metasurfaces and symmetry-broken plasmonic structures leverage structural non-centrosymmetry to achieve chiral light matter interactions[21]. Alternatively, intrinsically chiral materials-including chiral perovskites, chiral organic frameworks, and topological materials-exhibit inherent circular dichroism that enables CPL discrimination without requiring geometric asymmetry [14, 19]. Several comprehensive reviews have further summarized these developments, covering chiral metasurfaces, chiral halide perovskites, and chiral organic frameworks, and have highlighted the ongoing trade-offs among detection performance, operational bandwidth, and device integrability[6, 22-25].

Existing chiral metasurfaces often face a trade-off between high circular polarization discrimination and broadband tunability, motivating the development of compact, tunable chiral absorbers in the mid-infrared regime[26-28]. In this work, Here, we propose an asymmetric V-shaped plasmonic metasurface that supports chiral qBIC for selective CPL discrimination. By introducing in-plane symmetry breaking in a fixed-gap MIM configuration, a chiral qBIC resonance is induced with strongly spin-selective absorption: high under LCP while suppressed under RCP. The resonant wavelength is tuned from 5200 to 6200 nm via a uniform scaling factor S. Inspired by Wang et al.'s mirror-coupled qBIC absorber[11], our design fundamentally departs from their non-chiral approach by converting the polarization-independent BIC into a chiral qBIC with intrinsic spin selectivity.

## 2. Method

Numerical simulations were performed using the three-dimensional finite-difference time-domain (3D FDTD) method. Since the underlying gold film strictly blocks transmission (T ≈ 0), periodic boundary conditions and perfectly matched layers (PML) were applied in the *xy* and *z* directions, respectively, with a highly refined mesh within the structural region. LCP and RCP were individually and normally incident onto the device, with electric intensity of $E_0 = 1$ V/m. By extracting the reflection spectra $R$ and calculating the absorption using $A = 1-R$, the asymmetric absorption response and chiral selectivity of the device were revealed.

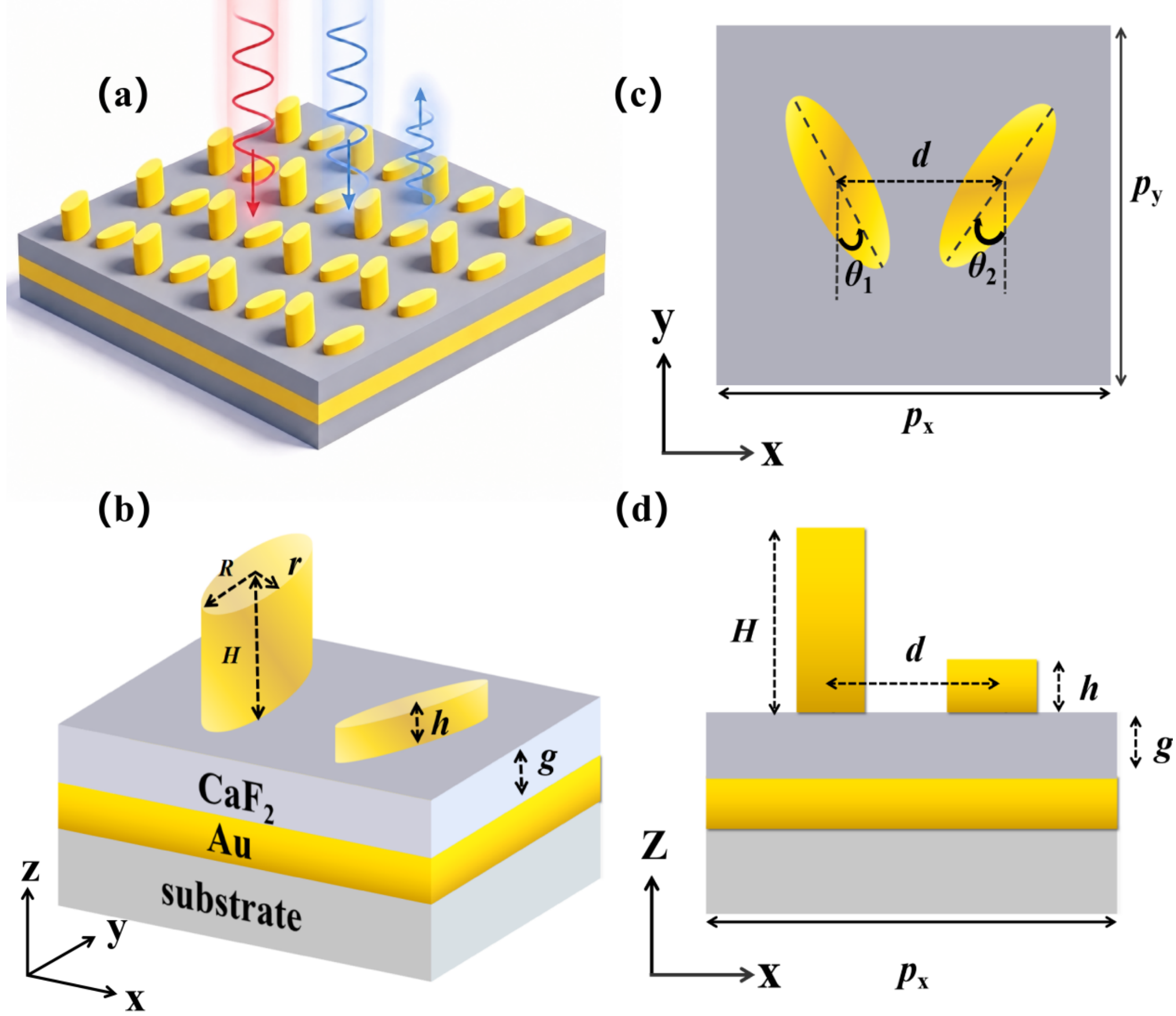

**Fig. 1.** Schematic of the device structure. (a) 3D view, with incident LCP and RCP; (b) 3D view of the unit cell, including the left elliptical cylinder (height $H$, semi-major axis radius $R$, semi-minor axis radius $r$) and the right elliptical cylinder (height $h$, semi-major axis radius $R$, semi-minor axis radius $r$), with the substrate being $CaF_2$ (thickness $g$) and Au layer (thickness 200 nm); (c) Top view (the $xy$ plane) of the unit cell, showing the angles between the major axis and y axis as $\theta_1$ and $\theta_2$ of the two ellipse, the distance between the centers of the two ellipse as $d$, and the periods in the $x$ and $y$ directions as $p_x$ and $p_y$ respectively; (d) Front view (the $xz$ plane) of the unit cell.

A three-dimensional schematic of the structure is shown in Fig.1. Define the top plane of calcium fluoride ($CaF_2$) as $z = 0$ plane, and the geometric center of the unit cell as the origin $(x,y) = (0,0)$ in the top view plane. The structure adopts a MIM vertical stacking configuration， consists from top to bottom, of gold (Au) asymmetric elliptical pair array, a $CaF_2$ dielectric spacer layer, a gold layer, which are on a $SiO_2$ substrate. The gold layer is sufficiently thick (larger than 200 nm), so no light is transmitted through it.

Inspired by the mirror-coupled plasmonic qBIC perfect absorber proposed by Wang et al., this study extends their symmetry-breaking parameter from the in-plane tilt angle to a three-dimensional geometry. By varying the height and tilt angle of the ellipse pairs, we obtain the optimal height and angle combination. The results show that this asymmetric design can effectively distinguish LCP and RCP, achieving high chiral discrimination.

**Table 1. Parameters of the Device**

| Parameters | $p_x$ | $p_y$ | $r$ | $R$ | $d$ | $h$ | $H$ | $g$ | $\theta_1$ (°) | $\theta_2$ (°) |
|---|---|---|---|---|---|---|---|---|---|---|
| unit: nm | 1600 | 1600 | 200 | 1000 | 1400 | 200 | 1500 | 250 | -4 | 20 |

## 3. Results and discussion

Fig.2 shows the absorption spectra of the symmetric and asymmetric structures under CPL incidence. For the asymmetric structure, the absorption spectrum for LCP light exhibits a distinct sharp resonance peak at approximately 5650 nm, with a peak intensity as high as 0.9. This resonance peak indicates the existence of a qBIC mode at resonant wavelength. In contrast, the absorption for RCP light is almost completely suppressed, reaching only 0.1 at the corresponding resonance and remaining below 0.1 across the measured spectral range. The pronounced absorption contrast demonstrates the chiral selectivity of the qBIC mode: near perfect absorption under LCP illumination versus strong suppression under RCP illumination, arising from the spin-dependent coupling efficiency of the symmetry-broken V-shaped structure.

To quantitatively describe this chiral discrimination capability, we introduce the circular polarization discrimination ratio ($DR$) for the absorption, which is defined as: $DR = 10 \times \log_{10} \frac{A_{\mathrm{LCP}}}{A_{\mathrm{RCP}}}$, where $A_{\mathrm{LCP}}$ and $A_{\mathrm{RCP}}$ correspond to the absorption intensity under LCP and RCP illumination, respectively. At the qBIC resonance peak ($A_{\mathrm{LCP}} \approx 0.9$, $A_{\mathrm{RCP}} \approx 0.1$), we obtain $DR = 8.6$. The high $DR$ indicates that the structure has excellent discrimination capability between LCP and RCP light at the qBIC resonance wavelength.

We also present the absorption spectra of the corresponding symmetric structure under CPL incidence. All the parameters of the symmetric structure are consistent with the ones of the asymmetric, except for $H = 200\,nm$ and $\theta_1 = 20°$. Due to the geometric symmetry, the absorption spectra for LCP and RCP light are identical, and thus both results are denoted as CPL. As shown by the black curve in the Fig.2, the absorption spectrum exhibits a resonance peak at approximately 5650 nm with a peak intensity of 0.72. Compared with the LCP resonance peak of the asymmetric structure, this peak shows a lower absorption value and a significantly larger full-width at half-maximum (FWHM). In summary, the broken-symmetry design of the asymmetric structure enables remarkable chiral discrimination capability.

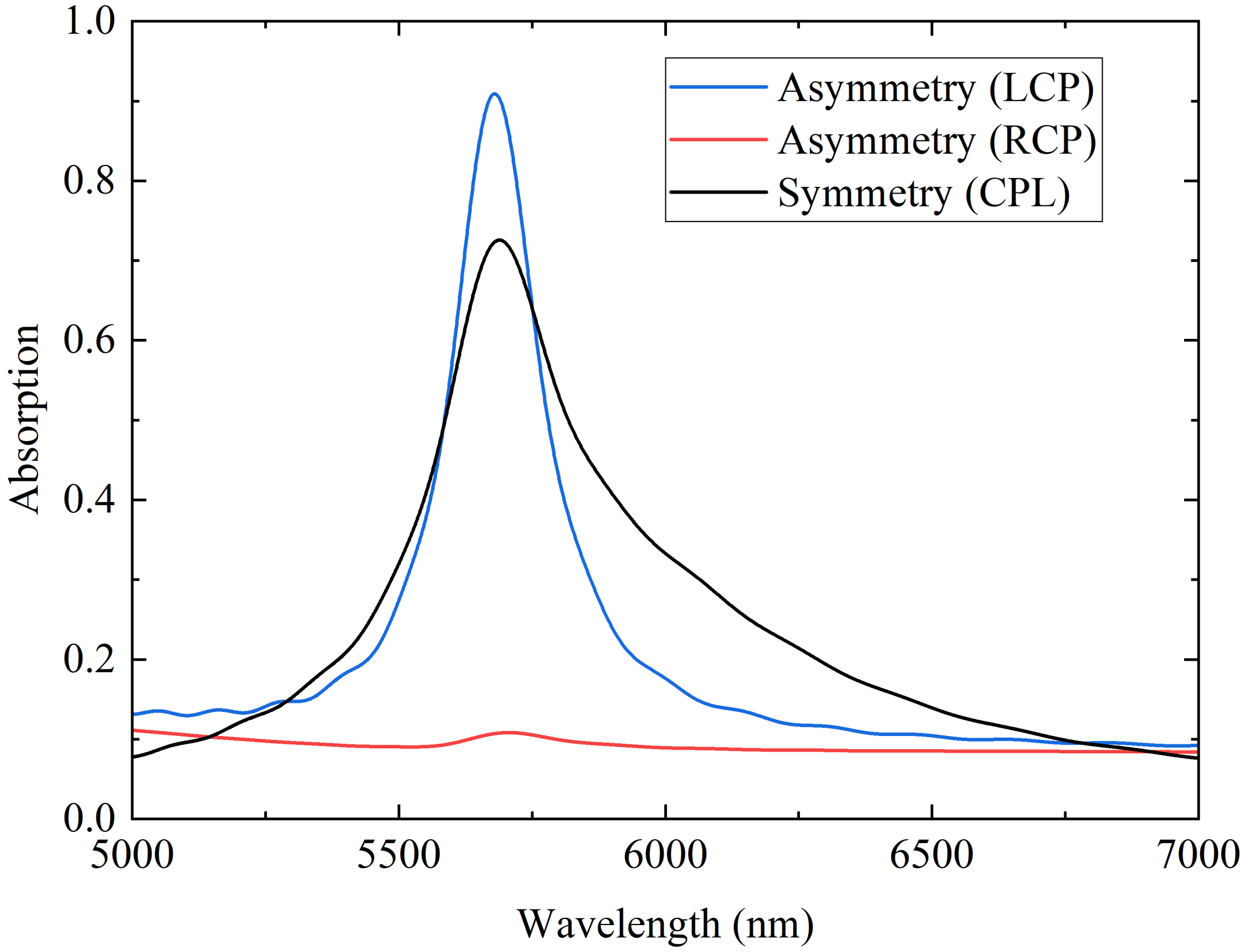

Fig.2 Absorption spectra. Blue and red curves represent the spectra under LCP and RCP incidence for the asymmetric structure. Black curve represents the absorption spectrum under CPL incidence for the symmetric structure.

Fig.3 presents the electric field distributions at three characteristic $z$ positions under LCP (upper row) and RCP (lower row) illumination at the resonant wavelength. At $z = -260$ nm (within the Au back layer), the field is weakest and broadly distributed, indicating minimal direct interaction with the lossy metal at this depth. Moving to $z = -125$ nm (within the $CaF_2$ spacer layer), the field becomes moderately enhanced and more concentrated around the resonator positions, yet remains relatively delocalized. The most pronounced field enhancement occurs at $z = 100$ nm (intersecting the top elliptical resonators), where the electric field is strongly localized near the tips of the elliptical cylinders, indicating efficient excitation of the qBIC mode.

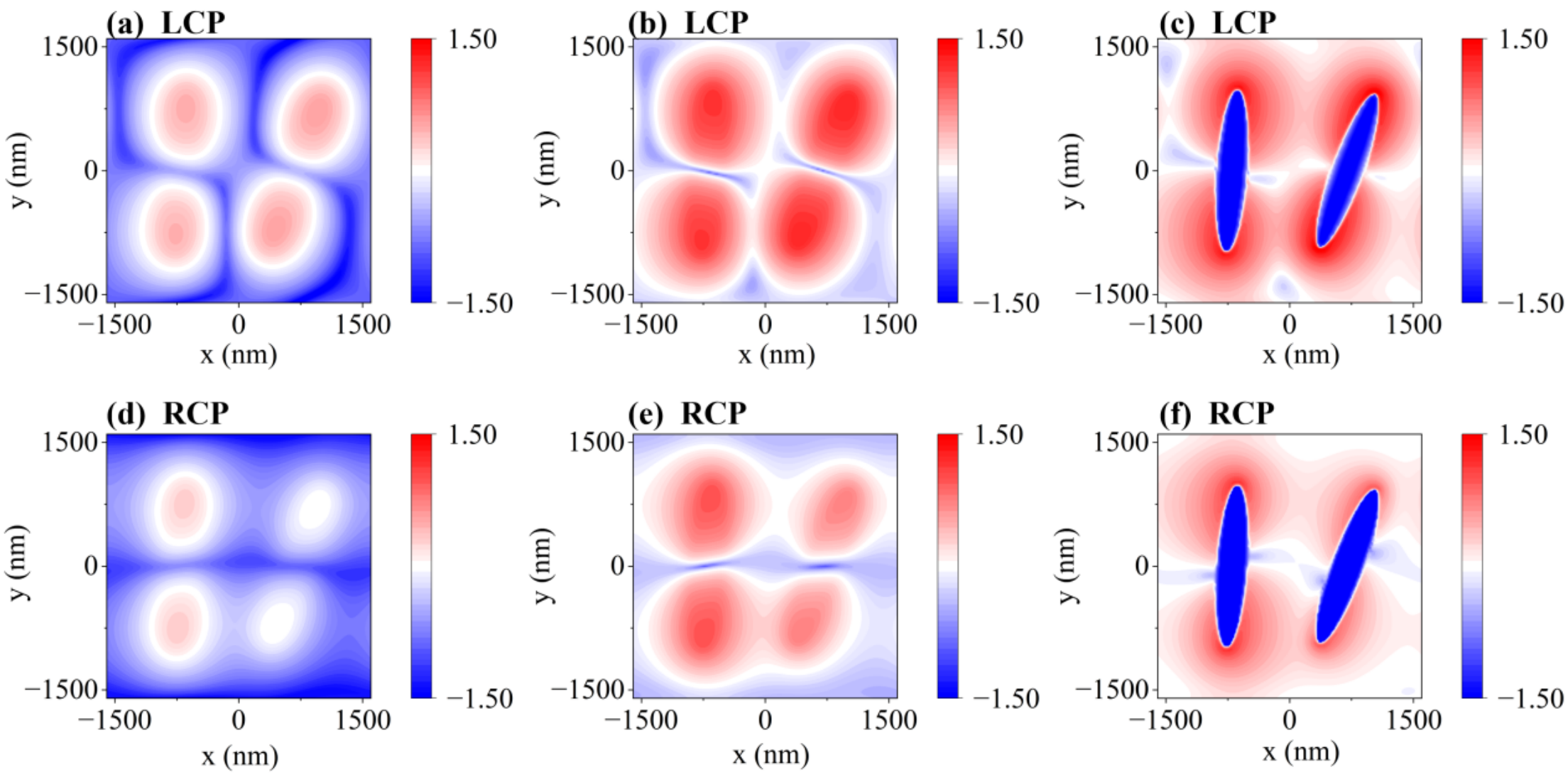

Fig.3 Electric field ($E$) distributions in the $xy$ plane at different $z$ positions under CPL for the asymmetric structure at the wavelength of 5678.08 nm. The upper and lower rows correspond to LCP and RCP illumination, respectively. The first, second, and third columns correspond to $z$=-260 nm (in the Au layer), -125 nm (in the $CaF_2$ layer), 100 nm (intersecting the elliptical cylinders) planes, respectively. Color bar represents $\log_{10}(E)$.

An excellent polarization-dependent asymmetry is observed at the resonator plane ($z = 100$ nm). Under LCP illumination, both elliptical arms exhibit intense and comparable field enhancement, with strong localization at the tips of each arm, indicating symmetric and efficient coupling of the LCP spin to the geometric chirality of the structure. In contrast, under RCP illumination, the overall field intensity is markedly weaker, and a pronounced left-right asymmetry emerges: the left arm shows noticeably stronger field confinement than the right arm. This spatial imbalance signifies incomplete and asymmetric excitation of the chiral qBIC mode under the mismatched spin state. At the other two planes ($z$ = -260, -125 nm), the electric field distributions are similar.

Fig.4 presents the electric field distributions in the $xz$-plane at three $y$-positions ($y$ = −500, 0, and 500 nm) under LCP (upper row) and RCP (lower row) illumination at the resonant wavelength. Across all cross-sections, the field is predominantly localized around the surfaces of the elliptical cylinders. At $y$ = −500 nm and $y$ = 500 nm, the field is primarily localized around the tips of the cylinder. On the other hand, at $y$ = 0 nm, the electric intensity remains much weaker. These results indicate the evident tip enhancement similar as shown in Fig.3.

A clear polarization-dependent spatial asymmetry emerges across the y-positions. At $y$ = −500 nm and $y$ = 500 nm, where the cross-sections intersect the upper tips of the elliptical cylinders, both sides exhibit pronounced field localization under LCP illumination, with strong tip enhancement visible around both the left and right cylinders. This symmetric bilateral confinement indicates that the LCP spin constructively couples to the geometric chirality of the structure, activating both arms of the asymmetric pair. In contrast, under RCP illumination at the same y-positions, the field distribution becomes strongly imbalanced: the left cylinder retains tip enhancement, whereas the right cylinder shows markedly weaker confinement. This left-dominant pattern under RCP indicates the destructive interference between the incident spin and the structural handedness, which selectively suppresses excitation on one side while permitting partial localization on the other. The reversal of the spatial asymmetry between LCP and RCP, i.e., symmetric bilateral tips versus left-dominant suppression, directly dominates the spin-selective nature of the chiral qBIC excitation.

The pronounced absorption contrast between LCP and RCP originates from the spin-dependent radiation loss rate of the chiral qBIC mode, as theoretically established by Overvig et al.[17] Under LCP illumination, the incident spin state matches the geometric handedness of the asymmetric elliptical pair, enabling destructive interference of the radiative emission from the two arms in the far field. This spin-matched configuration suppresses the radiative channel, yielding a high-quality qBIC resonance with narrow linewidth. When this resonance linewidth is comparable to the intrinsic material loss of the metallic resonators, the incident energy is efficiently confined and converted into heat, producing near-perfect absorption. The resulting coherent bilateral field enhancement at both cylinder tips, evident at $y$ = ±500 nm in Fig.4, confirms the excitation of a chiral qBIC resonance with coherent inter-arm coupling.

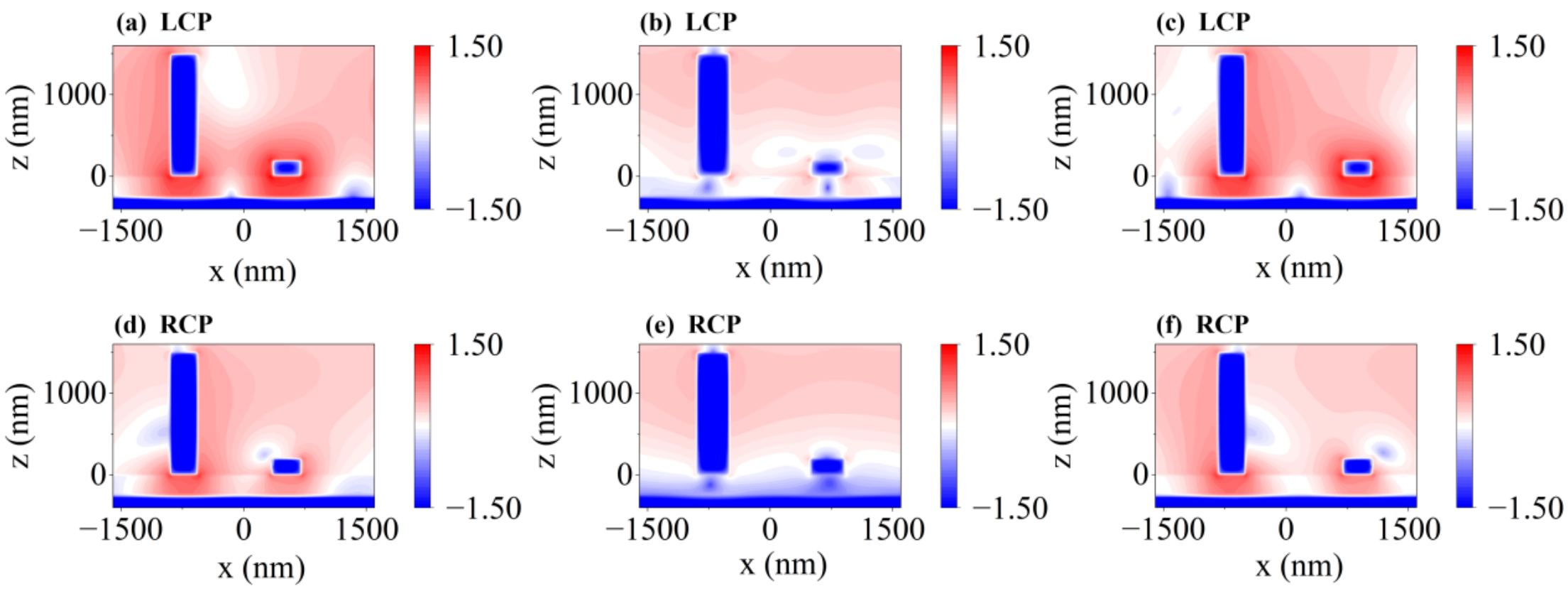

Fig.4 Electric field distributions in the *xz*-plane at different *y*-positions under circularly polarized light for the asymmetric structure at the wavelength of 5678.08 nm. The upper and lower rows correspond to LCP and RCP illumination, respectively. The first, second, and third columns correspond to cross-sections at $y$=−500 nm, $y$=0 nm, and $y$=500 nm, respectively. Color bar represents $\log_{10}(E)$.

Conversely, under RCP illumination, the mismatched spin state reverses the phase relationship between the two arms relative to the structural chirality, enhancing the effective asymmetry and opening the radiative channel, thereby substantially increasing the radiation loss rate. The resulting low quality factor precludes the formation of a long-lived resonant cavity mode. Although moderate field localization persists at the left cylinder tip, this represents an incoherent, single-arm near-field scattering state with large radiative damping rather than a collective qBIC resonance. Although the left cylinder tip exhibits moderate field localization under RCP illumination, this represents a short-lived, single-arm excitation that fails to establish coherent inter-arm coupling. Without the formation of a collective qBIC resonance, the incident energy cannot be effectively trapped within the structure for sufficient time to undergo Ohmic dissipation. Consequently, the RCP wave is largely reflected at the surface, yielding minimal absorption despite the localized field appearance.

This spin-selective transition, from coherent qBIC confinement under LCP to incoherent single-arm scattering under RCP, explains the large absorption contrast and high circular polarization discrimination of the proposed metasurface.

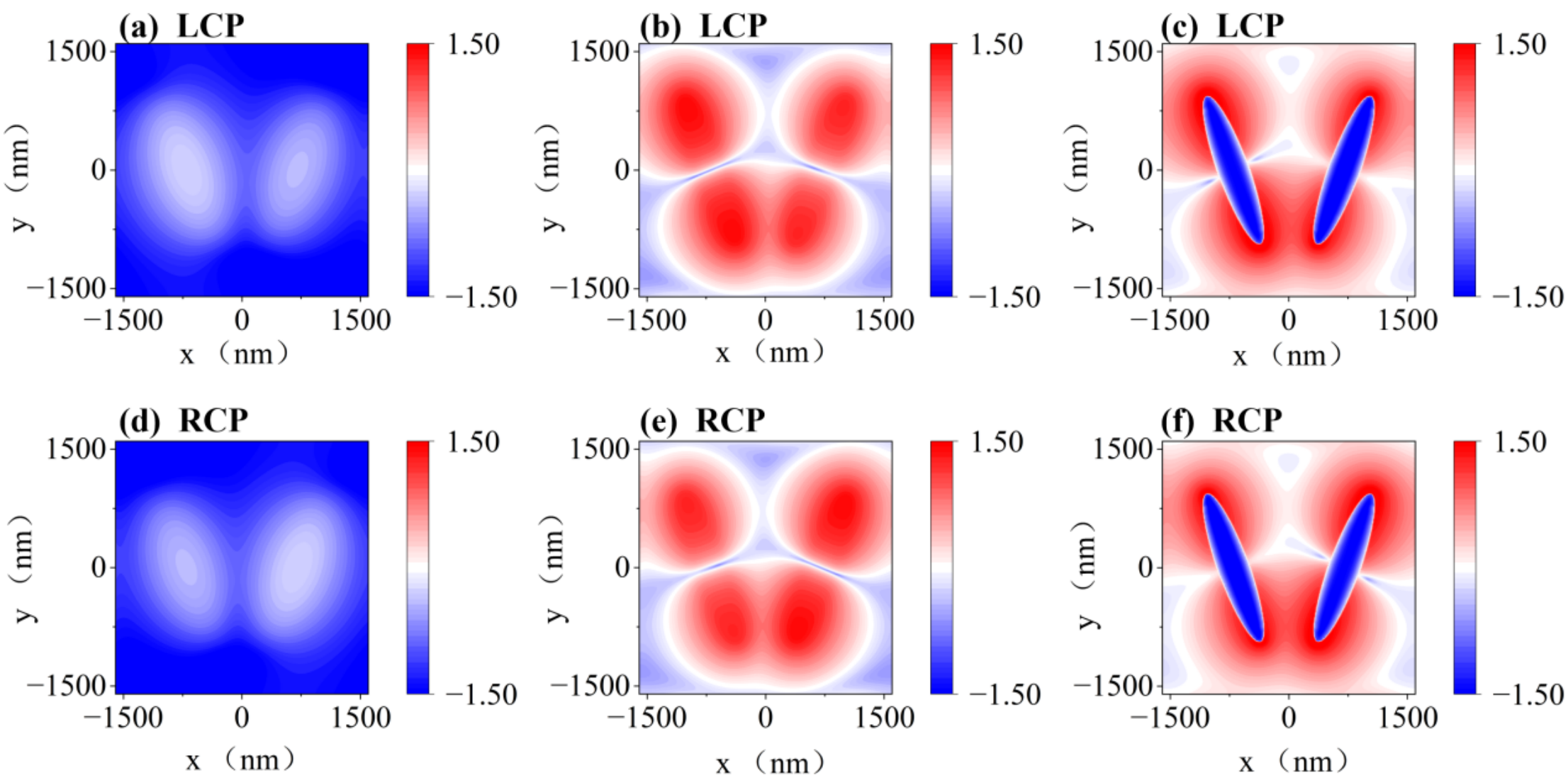

Fig.5 Electric field distributions in the *xy* plane at different *z* positions under CPL for the symmetric structure at the wavelength of 5689.35 nm. The upper and lower rows correspond to LCP and RCP illumination, respectively. The first, second, and third columns correspond to *z* = -260 nm (in the Au layer), -125 nm (in the $CaF_2$ layer), 100 nm (intersecting the elliptical cylinders) planes, respectively. Color bar represents $\log_{10}(E)$.

Fig.5 presents the *xy*-plane field distributions for the symmetric structure at the resonant wavelength. Unlike the asymmetric case (Fig.3), both LCP and RCP exhibit nearly identical field patterns at all *z* positions, with comparable intensity and spatial distribution. At the resonator plane (*z* = 100 nm), both polarizations show symmetric field enhancement around the elliptical cylinders, confirming the polarization-independent response of the mirror-symmetric configuration.

Fig.6 shows the corresponding *xz*-plane cross-sections. Again, LCP and RCP are almost indistinguishable, with field confinement symmetrically distributed around both cylinders and decaying uniformly into the $CaF_2$ spacer. No left-right asymmetry emerges for either spin state.

The absence of LCP-RCP differentiation in the symmetric structure confirms that the in-plane symmetry breaking is the sole origin of the chiral qBIC response. Without it, the structure supports only a polarization-independent resonance with moderate absorption (~0.7), lacking the near perfect spin-selective absorption enabled by the chiral qBIC mode in the asymmetric design.

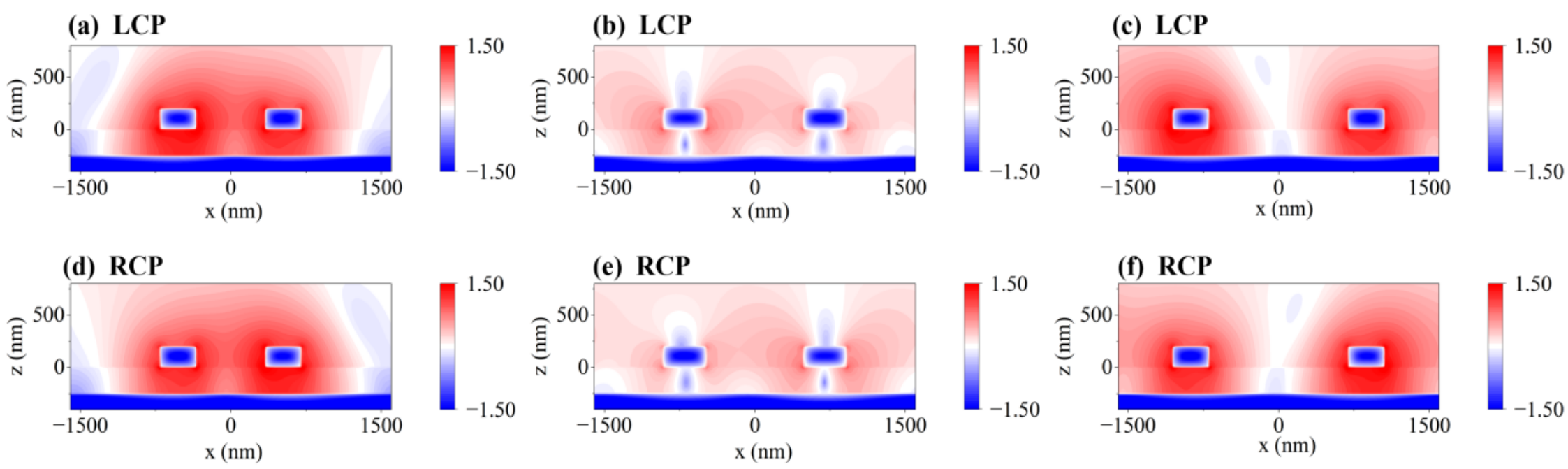

Fig.6 Electric field distributions in the *xz*-plane at different y-positions under CPL for the symmetric structure at the wavelength of 5689.35 nm. The upper and lower rows correspond to LCP and RCP illumination, respectively. The first, second, and third columns correspond to cross-sections at $y$=−500 nm, $y$=0 nm, and $y$=500 nm, respectively. Color bar represents $\log_{10}(E)$.

Fig.7 shows the LCP-RCP absorption difference spectra tuned by the scaling factor , i.e., $R' = S \times R$, $r' = S \times r$, $d' = S \times d$, $p_x' = S \times p_x$, $p_y' = S \times p_y$. As $S$ increases from 0.9 to 1.1, the resonance redshifts from about 5180 nm to about 6160 nm, consistent with the scaling invariance of qBIC resonances. The peak absorption difference exceeds 0.80 across the entire tuning range, indicating robust chiral selectivity. This broadband tunability, achieved without re-optimizing the structural asymmetry, offers a practical route for dynamic CPL detection and reconfigurable chiral devices.

Fig.8 quantifies the scaling robustness of the chiral qBIC response. Fig.8(a) shows that the resonance wavelength redshifts linearly from about 5180 nm to about 6160 nm as $S$ increases, consistent with geometric scaling. Meanwhile, the $DR$ remains above 8 dB across the entire range, indicating robust chiral selectivity. The right panel reveals that the LCP absorption peak stays near unity (~0.9) while the RCP peak remains below 0.15, confirming that the high absorption contrast is preserved under broadband tuning. The weak fluctuation in $DR$ (~1 dB) indicates minor deviations from optimal critical coupling at intermediate *DR* values, yet the overall performance remains excellent. These results demonstrate that uniform geometric scaling achieves predictable wavelength tuning without compromising the chiral qBIC discrimination, a key advantage for practical device implementation.

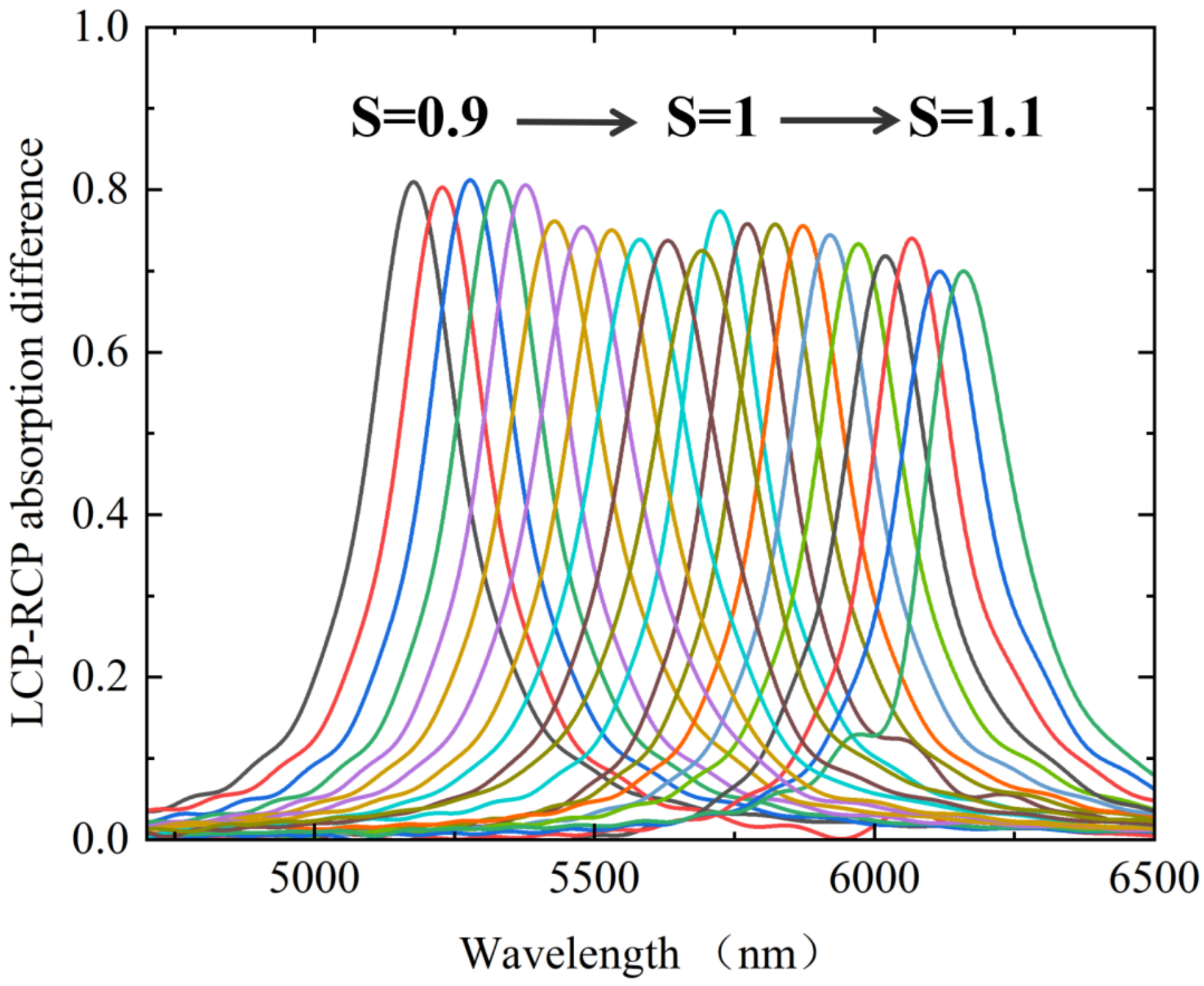

Figure 7. The evolution of the LCP-RCP absorption difference spectra with the scaling factor *S* varying from 0.9 to 1.1.

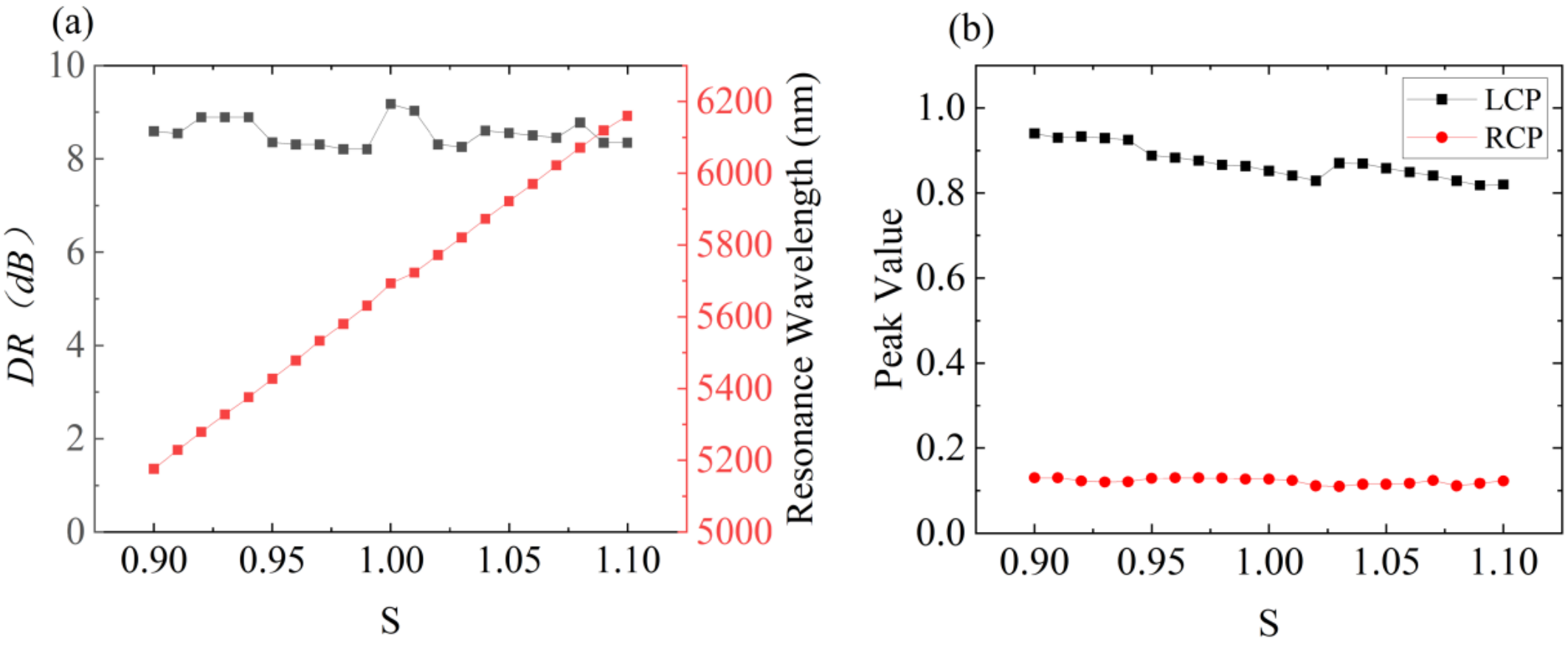

Fig.8 Influence of the scaling factor *S* on chiral absorption performance. (a) *DR* and resonance wavelength as functions of *S*. (b) LCP and RCP absorption peak values as a function of *S*.

## 4. Conclusion

In summary, we have demonstrated a tunable mid-infrared chiral selective absorber based on asymmetric V-shaped plasmonic metasurfaces driven by chiral qBICs. The structure achieves near perfect LCP absorption (about 0.9), strongly suppressed RCP absorption (<0.15), and a discrimination ratio exceeding 8 dB. Through uniform geometric scaling ($S$ = 0.9-1.1), the resonant wavelength is linearly tuned from 5200 to 6200 nm while maintaining high chiral selectivity ($DR > 8$ dB). This design offers a practical route for dynamic CPL detection and reconfigurable chiral nanophotonic devices.

## Acknowledgments

This work was supported by the National Natural Science Foundation of China (Grant No. 12504461).

## Disclosures

The authors declare no conflicts of interest.

## Data availability statement

The data that support the findings of this study are available upon reasonable request from the authors.